\documentstyle[preprint,aps,epsf,psfig]{revtex}
\tighten
\begin{document}

\title {Isoscalar compression modes in relativistic random phase
 approximation}

\author{Zhong-yu Ma$^{1}$\thanks{also Institute of Theoretical Physics,
    Beijing, P.R. of China}~, Nguyen Van Giai$^2$,
A. Wandelt$^{3}$, D. Vretenar$^{3}$\thanks{ On leave from University
  of Zagreb, Croatia} and P. Ring$^{3}$  \\
       1. China Institute of Atomic Energy, Beijing, P.R. of
     China\\
       2. Institut de Physique Nucl\'eaire, IN2P3-CNRS,
       F-91406 Orsay Cedex, France \\
       3. Physics Department, Technical
       University Munich, D-85748 Garching, Germany}

\date{22 March 2000}
\maketitle

\begin{abstract}
 A fully consistent relativistic RPA  calculation is performed for the
 monopole and dipole compression modes in nuclei. The emphasis is put on the
 effects of Dirac sea states which are generally neglected in relativistic
 RPA calculations. It is found that these effects can be quite important for
 the isoscalar monopole mode.
 The main contributions from the pairs of Fermi to Dirac sea
 states are through the exchange of the scalar meson, while the
 vector mesons play a negligible role. Numerical results of relativistic
 RPA are checked with the constrained relativistic mean field model in the
 monopole case. A good agreement beteween monopole energies calculated
 in RRPA and in time-dependent relativistic mean field approach is
 achieved. For the monopole compression mode, a comparison of
 experimental and calculated energies gives a value of 250 $\sim$ 270 MeV for
 the nuclear matter incompressibility. A large discrepancy remains
 between theory and experiment in the case of the dipole
 compression mode.

\vspace{2cm}

\end{abstract}


\section{introduction}

  The linear response of a system to a weak external field can be calculated
in the random phase approximation (RPA) which is one of the main
theoretical approaches to study the nuclear giant resonances in a
microscopic way\cite{ber75,rin80}. In the non-relativistic
approach
the equivalence between the RPA and the small amplitude limit of
the time-dependent Hartree-Fock approximation is well-known.
Numerous RPA calculations based on Hartree-Fock single-particle spectra
have been performed
using effective nucleon-nucleon interactions such as Skyrme and
Gogny forces. These calculations are very successful in describing
the collective excitations and giant resonances in various nuclei
in a framework which is consistent in the sense that both mean
field and residual interaction come from the same effective force.

  In recent years the relativistic mean-field (RMF) theory with non-linear
effective Lagrangians has received considerable attention. The RMF
has met great success in describing the bulk properties of nuclei.
One of the advantages of the RMF approach is that the same
effective Lagrangian with only a few free parameters could predict
the ground state properties of nuclei covering the whole periodic
table and also exotic nuclei from the proton drip line to the
neutron drip line.  The success of the RMF encouraged to
investigate the extensions of the RMF approach to study
dynamical processes, especially the collective giant resonances.
The linear response of a system to an external field can be
calculated in the relativistic RPA (RRPA) which is again a small
amplitude limit of the time-dependent RMF (TDRMF)
theory\cite{rin99}.

 In the relativistic approach, it is known from nuclear matter studies
that self-consistency is extremely important.
Self-consistent wave functions where the lower components in the
Dirac spinor are enhanced by an effective mass must be adopted in
the Dirac-Brueckner Hartree-Fock calculations\cite{bro90}. Two
questions of consistency appear in the RRPA. First, the
particle-hole interaction must be determined from the same
Lagrangian used in the RMF.
The meson nonlinear self-interaction terms in the effective
Lagrangian bring in density-dependent interactions which are
essential for a quantitative description of the nuclear
properties, especially the nuclear matter compression modulus. The
meson propagators with nonlinear terms have been worked out in
Refs.\cite{mapr97,manp97}. The second question of consistency is
that the RRPA  must be consistent with the RMF. It is assumed in
the RMF that only positive energy states contribute to the nucleon
self-energies (no-sea approximation).
A complete relativistic Hartree approximation would have to
include the contributions from Dirac states, which are divergent
and a renormalization procedure is required\cite{chi77}.
Unfortunately, a proper treatment of the relativistic Hartree
approach in finite nuclei is too complicated and one has to work
in the no-sea approximation. Nevertheless, a consistent RRPA built
on an RMF ground state must deal in principle with the Dirac
states. This is because RRPA is a small amplitude limit of TDRMF
and the density change $\delta \rho (t)$ at time $t$ has
necessarily matrix elements not only between positive energy
particle-hole pairs but also between pairs formed from Dirac sea
states and occupied Fermi sea states of the static solution.
Thus, an RRPA calculation consistent with the no-sea approximation
should include both positive energy and negative energy
single-particle states.
Although it has been known before\cite{daw90} that the inclusion
of the Dirac states ensures current conservation and decoupling of
the spurious state, it was thought that the contribution to the
RRPA energies and strengths from the Dirac states would be
negligible.  Therefore, the observables of interest (excitation
energies and transition densities) would remain the same with and
without including the Dirac states. Most previous works in
studying the response functions of nuclei included the positive
energy particle-hole pairs only. With this prescription,
discrepancies between the RRPA and the TDRMF appeared, especially
in the isoscalar monopole modes, which could not be understood for
some time\cite{nvg99,vre99}. The main purpose of this paper is to
clarify this puzzle and to examine the effects of the Dirac states
on the calculation of compression modes in the RRPA approach.
We analyze the origin of the contributions from Dirac states in
the study of the isoscalar nuclear giant resonances. To verify
that the calculation of RRPA with Dirac states is the correct and
consistent  procedure we compare the inverse energy-weighted sum
rules calculated in RRPA and in constrained RMF calculations. A
good agreement can be obtained only when Dirac states are
included. Results concerning isoscalar giant monopole resonances
(ISGMR) and isoscalar giant dipole resonances (ISGDR) in nuclei
are presented.

The outline of this paper is as follows. The formalism of the
relativistic linear response is outlined in Sec.II where we
investigate the implications of including Dirac states in the RRPA
equations. In Sec.III the constrained RMF method is used to
calculate isoscalar monopole polarizabilities which are compared
to those deduced from RRPA.
Results of calculations of ISGMR and ISGDR in medium and heavy
nuclei are presented in Sec.IV. Finally, Sec.V contains a summary
of the main results.

\section{Dirac states in the RRPA}

  We start from the Hartree single-particle Green's function
which  is defined by:
\begin{equation}
G_H(x,y)=-i\langle 0|T(\Psi_H(x)\overline{\Psi}_H(y))|0 \rangle~,
\label{eq1}
\end{equation}
where $x \equiv (t, {\bf r})$, $\Psi_H(x)$ is the Hartree field
operator and $|0\rangle$ is the uncorrelated Hartree ground state.
Then, the Dyson equation for the nucleon propagator is
\begin{equation}
G_H(x,y)=G_0(x,y)+\int d^4x_1 d^4x_2 G_0(x,x_1)
  \Sigma^H(x_1,x_2)G_H(x_2,y)~,
 \label{eq2}
 \end{equation}
where the nucleon self-energy due to meson-nucleon couplings can
be calculated in the Hartree approximation including only the
tadpole Feynman diagram:
\begin{eqnarray}
\Sigma^H(x,y)& = & \delta^4(x-y) \{-ig^2_s \int d^4x_1
D_\sigma(x-x_1) Tr [G_H(x_1,x_1^+) ]  \nonumber \\
 &&-i\gamma_\mu g^2_v \int d^4 x_1 D_\omega^{\mu \nu}(x-x_1)
  Tr [ \gamma_\nu G_H(x_1,x_1^+)] \} \nonumber \\
 &=&\delta^4(x-y)[\Sigma_s(x)-\gamma_\mu \Sigma_v^\mu(x)]~,
\label{eq3}
\end{eqnarray}
where $x^+$ refers to the time-ordering prescription and
$D_\sigma$, $D_\omega$ are the meson propagators of $\sigma$ and
$\omega$, respectively. For the convenience of notations, we write
explicitly only the $\sigma$ and $\omega$ mesons in this section
but the actual calculations of the next sections contain also the
$\rho$ meson and photon fields.

The Hartree Green's function in a standard spectral representation
is expressed as:
\begin{eqnarray}
G_H({\bf r}_1,{\bf r}_2;E)&=& \sum_\alpha \frac{\phi _\alpha({\bf
r}_1) \overline{\phi}_\alpha({\bf r}_2)}
 {E-\varepsilon_\alpha+i\eta}
+\sum_{\bar{\alpha}} \frac{\phi _{\bar{\alpha}}({\bf r}_1)
\overline{\phi}_{\bar{\alpha}}({\bf r}_2)}
{E-\varepsilon_{\bar{\alpha}}-i\eta}\\ \nonumber
  &&+\sum_{h<\varepsilon_F} \phi _h({\bf r}_1) \overline{\phi}_h({\bf r}_2)
 \left(\frac{1}{E-\varepsilon_h-i\eta} -\frac{1}{E-\varepsilon_h+i\eta}\right)
 \\ \nonumber
 &\equiv & G^F_H+ G^D_H~,
 \label{eq4}
 \end{eqnarray}
where we denote by $\alpha$ all positive energy states
($\varepsilon_\alpha >0$), $\bar{\alpha}$ all negative energy
states ($\varepsilon_{\bar{\alpha}}<0$) and $h$ the states in the
Fermi sea. The unoccupied positive energy states will be denoted
by $p$ in the following expressions. The superscripts $F$ and $D$
refer to the Feynman and density-dependent parts of the Green's
function, respectively. When one calculates the nucleon
self-energy of Eq.(3) with this Green's function Eq.(4), all
occupied states $\bar{\alpha}$ and $h$ will contribute. The
contributions from occupied Dirac states are divergent and
therefore, a renormalization procedure is necessary. This
corresponds to the so-called relativistic Hartree approximation.
In contrast, the RMF approach which is the framework we adopt here
assumes no contributions to the nucleon self-energies from the
Dirac states. This no-sea approximation corresponds to a modified
single-particle Green's function, where the poles
$\varepsilon_{\bar{\alpha}}+i\eta$ for negative energy states are
shifted to $\varepsilon_{\bar{\alpha}}-i\eta$ in
Eq.(4)\cite{daw90}. This gives a modified Green's function
$\tilde{G}_H$ built on a complete set of single-particle states
and it satisfies the Dyson equation Eq.(\ref{eq2}). A consistent
RRPA built on the RMF ground state has to start from this modified
single-particle Green's function.

The RRPA method used in this paper is described in detail in
Refs.\cite{lhu89,mapr97} with just the replacement of ${G}_H$ by
$\tilde{G}_H$. The linear response of a system to an external
field is given by the imaginary part of the retarded polarization
operator:
\begin{equation}
R(Q,Q;{\bf k},E)=\frac{1}{\pi} Im \Pi^R(Q,Q;{\bf k},{\bf k};E)~,
\label{eq5}
\end{equation}
where Q is a one-body operator represented by a 4$\times$4 matrix.
The polarization operator can be obtained by solving the
Bethe-Salpeter equation:
\begin{eqnarray}
&&\Pi(P,Q;{\bf k},{\bf k'},E)=\Pi_0(P,Q;{\bf k},{\bf k'},E) \\
\nonumber
&& -\sum_{i} g_i ^2 \int d^3 k_1 d^3 k_2 \Pi_0(P,\Gamma
^i;{\bf k},{\bf k}_1,E)
 D_i ({\bf k}_1-{\bf k}_2, E)
 \Pi(\Gamma_i,Q;{\bf k}_2,{\bf k'},E)~.
\label{eq6}
\end{eqnarray}
In this equation the index $i$ runs over $\sigma$, $\omega$ and
$\rho$ mesons, $g_i$ and $D_i$ are the corresponding coupling
constants and meson propagators. The meson propagators for
non-linear models are non-local in momentum space and they have to
be calculated numerically. The detailed expressions of the $D_i
({\bf k}_1-{\bf k}_2, E)$ can be found in
Refs.\cite{mapr97,manp97}. The $\Gamma_i$'s are the 4$\times$4
matrices 1, $\gamma_{\mu}$ and $\gamma_{\mu}\vec{ \tau}$ for
$i=\sigma , \omega$ and $\rho$, respectively. Notice that the
space-like parts of vector mesons play no role in the static ground
states, but they provide attractive contributions to the
particle-hole residual interaction.
The kernel of the
Bethe-Salpeter equation has a simple form in the RMF
approximation:
\begin{eqnarray}
\Pi_0(P,Q;x_1,x_2)&=&i~ \langle 0|T [ \overline{\Psi}_H(x_1)P
\Psi_H(x_1) \overline{\Psi}_H(x_2)Q \Psi_H(x_2) ]|0\rangle \\
\nonumber &=&i~Tr [P\tilde{G}_H(x_1,x_2)Q\tilde{G}_H(x_2,x_1)]~.
\label{eq7}
\end{eqnarray}

In the spectral representation, the unperturbed polarization
operator has the following retarded form:
\begin{equation}
\Pi_0^R(P,Q;{\bf k},{\bf k'},E)= \sum_{h,~a=p,\bar{\alpha}} \left[
 \frac{\langle \overline{\Psi}_h|P| \Psi_a \rangle
      \langle \overline{\Psi}_a|Q| \Psi_h \rangle}
     {E-(\varepsilon_a-\varepsilon_h)+i\eta}
 +\frac{\langle \overline{\Psi_a}|P| \Psi_h \rangle
      \langle \overline{\Psi_h}|Q| \Psi_a \rangle}
     {E+(\varepsilon_a-\varepsilon_h)+i\eta} \right],
\label{eq8}
\end{equation}
which shows that the unperturbed polarization includes not only
particle-hole pairs but also pairs formed from the Dirac and Fermi
states $\bar{\alpha}h$, when one substitutes $\tilde{G}_H$ in
Eq.(7). The terms with $\bar{\alpha}h$ pairs would correspond to the
Pauli-violating terms which appear if one works with $G_H$ in
Eq.(7)\cite{lhu89}. There is no Pauli violation in the present approach
because we work with the modified Green's function $\tilde{G}_H$.

   In the time-dependent RMF (TDRMF) approach
the no-sea approximation is imposed at each time. In the small
amplitude limit the density $\rho(t)$ can be expanded as the
density at $t=0$  and a small density change $\delta\rho(t)$
which can be calculated in a complete set of states given by the
static solution of RMF. This complete set includes the positive as
well as negative energy states.
Thus, $\delta\rho$ has non-zero matrix elements $\delta\rho_{ph}$
and $\delta\rho_{\bar{\alpha}h}$.
Therefore, the particle-hole space of RRPA
must be built with both positive and negative  energy states.
The equivalence between the RRPA including the Dirac states and
the small amplitude limit of the TDRMF with no-sea approximation
can be formally shown\cite{rin99}.

In Eq.(8), one finds that the particle-hole energies between the
occupied positive energy states and the Dirac states are usually
two orders of magnitude larger than those
%of the particle-hole states
involving only positive energy states. Furthermore, the matrix
elements of $P$ and $Q$ between positive and negative energy
states should be small because the positive (negative) energy
states have a small lower (upper) component  in their Dirac
spinors. It is thus found that the $\bar{\alpha}h$ pairs make no
visible contribution to the imaginary or to the real part of the
unperturbed polarization in the energy region of the
nuclear collective excitations. This is the reason why the Dirac
states were usually not included in most previous works of the
RRPA. However, we will see in the next sections that the
correlated polarization operator can be affected by the Dirac
states when the particle-hole interaction depends on the $\sigma$
meson, i.e., in the case of isoscalar excitations.

\section{Constrained RMF}

In the non-relativistic approach there is a complete equivalence
between the static polarizability calculated in RPA and that
deduced from a constrained Hartree-Fock calculation, and this
property can be used as a check of consistency of the
calculations. The same property will be used here to compare the
static polarizabilities of RRPA and constrained RMF and to
investigate the importance of the inclusion of Dirac states in the
RRPA.
The constrained RMF method was used in Ref.\cite{mar89} to
evaluate the ISGMR energies within the Walecka's linear model. In
this section we discuss the constrained RMF results obtained with
effective lagrangians having different types of self-interactions
in the $\sigma$ and $\omega$ fields.

  In general, a constrained RMF equation with an external field $\lambda Q$ is
described as:
  \begin{equation}
  \left(H(\lambda)+\lambda Q \right)|\lambda\rangle =E(\lambda)|\lambda
  \rangle~,
\label{eq9}
\end{equation}
where $H(\lambda)$ is the RMF hamiltonian, which depends on the
parameter $\lambda$ due to the self-consistency of the
single-particle states, $|\lambda\rangle$ and$E(\lambda)$ are the
ground state and corresponding energy of
$H(\lambda)+\lambda Q$, respectively. At $\lambda$=0 the first
order variation of $\langle \lambda \vert Q \vert \lambda \rangle$
with respect to $\lambda$, i.e., the static polarizability is
equal to the second order variation of $E(\lambda)$. The same
static polarizability can be calculated as twice the inverse
energy-weighted sum rule $m_{-1}$ of RRPA,
\begin{eqnarray}
\frac{\partial \langle \lambda|Q|\lambda\rangle}{\partial \lambda}
\vert_{\lambda=0} & = & \frac {\partial ^2
E_{\lambda}}{\partial\lambda^2} \vert_{\lambda=0} \nonumber \\
 & = & 2 \sum_{n\ne 0}\frac{\langle 0|Q|\Psi_n \rangle \langle
 \Psi_n|Q|0\rangle}  {E_n-E_0}~.
\label{eq10}
\end{eqnarray}
The second line of Eq.(\ref{eq10}) shows that the polarizability
can be deduced from the real part of $\Pi$ calculated at $E$=0. In
the case of the ISGMR, the operator $Q$ is simply $\gamma_0 r^2$ and the
RRPA $m_{-1}$ sum rule can also be obtained by performing a
spherically constrained RMF calculation. Note that, if one
replaces in Eq.(\ref{eq9}) the self-consistent hamiltonian
$H(\lambda)$ by the fixed hamiltonian $H_0$ of the unconstrained
RMF solution, then Eq.(\ref{eq10}) gives the $m_{-1}$ sum rule of
the unperturbed response corresponding to the polarization
operator $\Pi_0$.
%The expectation of $Q$ is $A <r>^2=\langle 0|r^2|0\rangle$. The
%unperturbed EIWS corresponds to the case of fixed Hamiltonian $H(0)$.

   The effective Lagrangians we
use here include the scalar meson $\sigma$, vector meson $\omega$,
isovector vector meson $\rho$ and the photon field $A^{\mu}$.
Self-interactions in the $\sigma$ and $\omega$ fields are
introduced in order to simulate a density dependence and hence to
lower the compression modulus of nuclear matter\cite{Boguta}:
\begin{eqnarray}
 {\cal L}&=&\overline{\Psi}(i\gamma^{\mu}\partial_{\mu}-M_N-g_\sigma
\sigma -g_\omega \gamma^\mu \omega_\mu
          -g_\rho \tau^a\gamma^\mu \rho^a_\mu
      -e\gamma^\mu A_\mu \frac{1}{2}(1-\tau_3))\Psi  \\ \nonumber
  && +\frac{1}{2}\partial^\mu \sigma \partial_\mu \sigma
      -U_{\sigma}  -\frac{1}{4}W^{\mu \nu} W_{\mu \nu}+ U_{\omega}
       +\frac{1}{2}m^2_\rho \rho^{a \mu} \rho^a_{\mu}
      -\frac{1}{4}R^{a \mu \nu} R^a_{\mu \nu}
     -\frac{1}{4}F^{\mu \nu} F_{\mu \nu}~,
\label{eq11}
\end{eqnarray}
where
\begin{eqnarray}
W^{\mu \nu}&=&\partial ^\mu \omega^\nu - \partial ^\nu
\omega^\mu~, \\ \nonumber
 R^{a\mu \nu}&=&\partial ^{\mu}\rho ^{a \nu} - \partial ^{\mu}\rho ^{a \nu}
               +g_\rho \epsilon^{abc}\rho^{b\mu}\rho ^{c \nu}~, \\ \nonumber
F^{\mu \nu}&=&\partial ^\mu A^\nu - \partial ^\nu A^\mu~,
\label{eq12}
\end{eqnarray}
\begin{eqnarray}
 U_{\sigma}= \frac{1}{2}m_\sigma ^2 \sigma^2 +\frac{1}{3}g_2\sigma ^3
              +\frac{1}{4}g_3 \sigma^4, ~~
   U_{\omega}=\frac{1}{2}m^2_\omega \omega ^\mu \omega _\mu
    +\frac{1}{4}c_3(\omega ^\mu \omega _\mu)^2~.
\label{eq13}
\end{eqnarray}
We briefly outline the procedure of the numerical calculations.
The nucleon mean fields are calculated by solving the Dirac-Hartree
equation in coordinate space self-consistently in the RMF
approximation. All single-particle states (unoccupied and occupied) are
calculated
by diagonalizing the self-consistent mean fields in a harmonic
oscillator basis with a proper choice of the oscillator length $b$
and the number of shells ($N$=12). An averaging parameter
$\Delta$=2MeV is introduced as the imaginary part of the energy to
smooth out the response functions. Calculations are performed with
different non-linear lagrangians which span a wide range of values
of the nuclear incompressibility:
NL1\cite{rhe86}, NLSH\cite{sha93}, NL3\cite{lal97} and
TM1\cite{sug94}. They generally give good descriptions of the
ground state properties.  The linear Walecka model HS\cite{hor81}
is also considered for comparison because of its high
incompressibility. In the constrained RMF calculations we use a
small step $\Delta \lambda$=0.5$\times$10$^{-4}$fm$^{-3}$ in order
to calculate the derivatives with sufficient accuracy.

The values of the inverse energy-weighted sum rules calculated in
constrained RMF and in RRPA are compared in Table 1 for the case
of ISGMR in $^{208}$Pb.
We find a  good agreement(within 1\%) between the first two
columns, which is a numerical check of the accuracy of the
constrained RMF calculations.
The values of $m_{-1}$ in the third column are calculated by
integrating the inverse energy-weighted ISGMR strength (from 0 to
60 MeV).
The last column corresponds to one half of the real part of the
response function at zero energy. The agreement between the last
two columns simply indicates that the dispersion relation between
the real and imaginary parts of $\Pi$ or $\Pi_0$ is well obeyed
numerically.
The comparison between the constrained RMF and the RRPA provides a
severe check for the RRPA method used in this paper. It is found
that the agreement is quite good (within 1\%) for the unperturbed
cases while the difference becomes a few percent for the RRPA
cases.
We find that the discrepancies between RRPA and constrained RMF
are
mainly due to the Coulomb force. If one switches off the Coulomb
force the agreement becomes perfect (less than 1\% in all cases).
The expansion of the wave functions on the harmonic oscillator
basis may not be accurate enough to account for the long range
effects of the Coulomb force
especially for the Dirac states with an attractive Coulomb
interaction. A more sophisticated method would be to calculate the
Green's functions directly in coordinate space\cite{wehr}.

The discrepancies between the RRPA and the constrained RMF are
dramatic if one does not take the Dirac states into account as one
can see in row III of Table  1.
The large contribution to the RRPA strengths from the Dirac states
can be understood in the following way. In the relativistic
approach the nucleon potential  is obtained by a strong
cancellation between the isoscalar scalar and vector potentials,
and so do the particle-hole residual interactions in the isoscalar
channel. Due to the  $\gamma$-matrix structure the matrix elements of the
residual interaction between particle-hole pairs through the
exchange of scalar and vector mesons are largely cancelled to two
orders of magnitude. However, this is not true for pairs built
with Fermi and Dirac states because the large component of the
wave function is the upper component (lower component) for a Fermi
(Dirac) state.
The orthogonality between positive and negative energy states with
the same quantum numbers is mainly  due to the cancellation
between the upper and lower components. Again, due to the
$\gamma$-matrix structure
the matrix elements between Fermi and Dirac states through the
exchange of vector mesons ($\omega$ or $\rho$) are small whereas those
corresponding to a scalar meson are large. These scalar meson
matrix elements are
repulsive and they can overcome the large energy gap between Dirac
and Fermi states.

\begin{figure}[ht]
%\centerline{\psfig{figure=fig1.eps,height=14cm,width=14cm.angle=90}}
\epsfxsize=\hsize\epsfbox{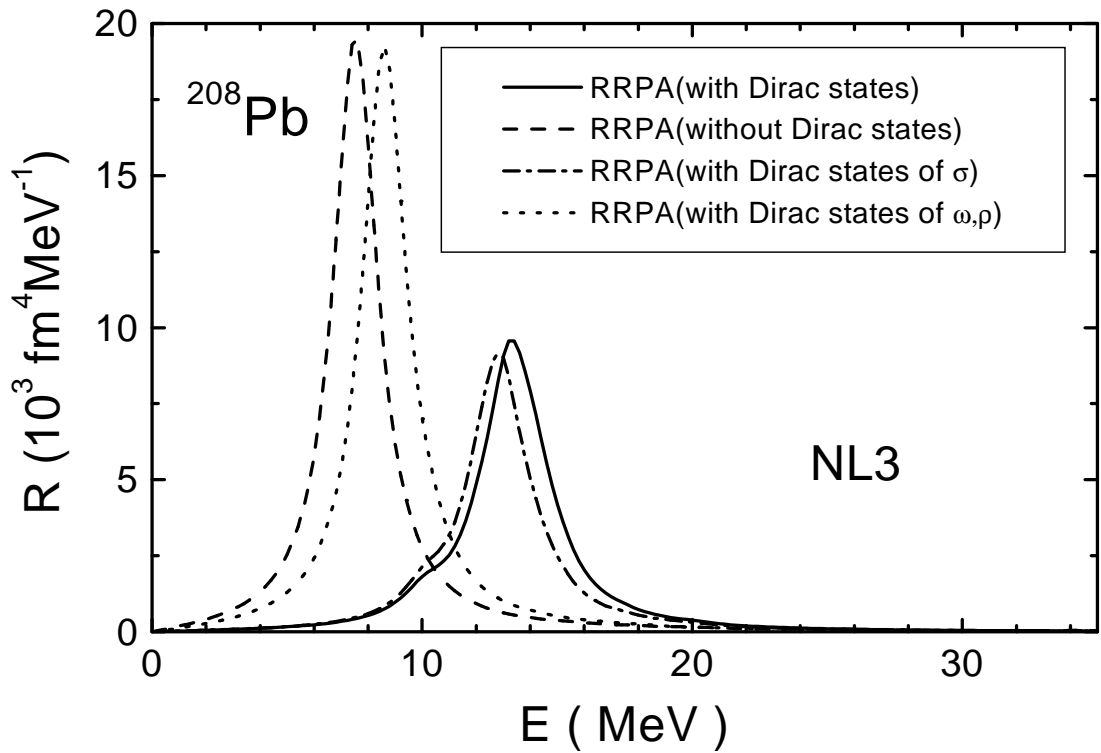}
%\vglue-0.5cm
\caption{ ISGMR strength distributions in $^{208}$Pb calculated with NL3
parametrization. The solid and long-dashed curves are the RRPA
strengths with and without Dirac states, respectively.  The
short-dashed (dash-dotted) curve corresponds to calculations where
only vector mesons (scalar mesons) are included in the couplings
between Fermi and Dirac states.}
\end{figure}

The large effects of Dirac states are illustrated in Fig.1. We
separate the contributions from scalar and vector mesons for the
$\bar{\alpha}h$ pairs and we plot the monopole strengths in
$^{208}$Pb, taking NL3 as an example. The solid and long-dashed
curves are the RRPA strengths with and without Dirac states,
respectively. They are markedly separated. The ISGMR strength is
pushed strongly down without the Dirac state contributions.
The short-dashed curve is the RRPA strength with the contributions
of $\bar{\alpha}h$ pairs coming from the vector meson only.
It is more or less the same as the case without Dirac states. The
dash-dotted curve is calculated with only the scalar meson
contributions to the $\bar{\alpha}h$ pairs.
It is clearly seen that the contributions of the scalar meson in
the $\bar{\alpha}h$ pairs dominate whereas those due to the vector
meson are negligible.

It can be concluded that the inclusion of only the positive
particle-hole pairs in the isoscalar modes, where the isoscalar
mesons play an important role, would provide a too strong
attraction and therefore, a too low isoscalar giant resonance
energy. This will reflect in a too large polarizability as shown
in Table 1.
On the other hand, the contributions from Dirac states can be
neglected in the case of isovector modes where the
isovector-vector meson dominates.

\section{Monopole compression mode}
The value of the compression modulus $K_\infty$ of nuclear matter
is one of the most important
issues of nuclear physics. The generally accepted procedure for
determining $K_\infty$ is to calculate isoscalar monopole
energies, as well as other properties of finite nuclei using
various microscopic models in order to select those which can
describe correctly the data and to use these models for
constructing the nuclear matter equation of state\cite{bla80}.
Here, we examine the predictions of the various effective
lagrangians in the framework of RRPA. In this work we adopt the
mean field approach and therefore the effects of 2p-2h
couplings\cite{colo99} are not included.

\begin{figure}[htb]
%\centerline{\psfig{figure=fig2.eps,height=14cm,width=14cm.angle=90}}
\epsfxsize=\hsize\epsfbox{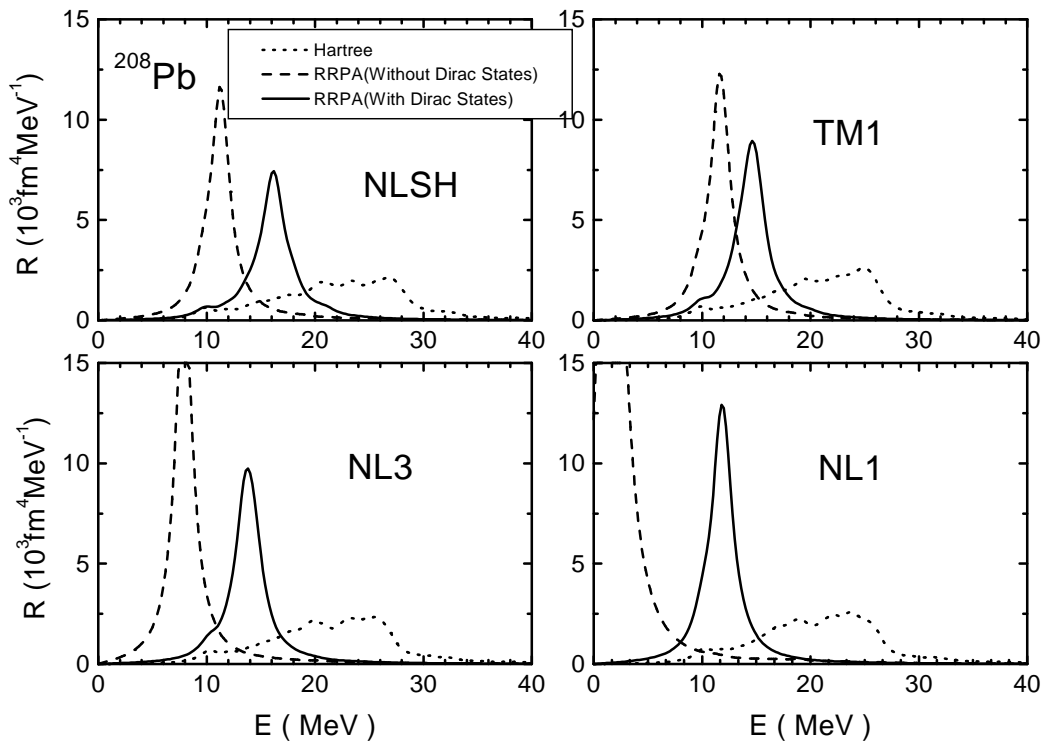}
%\vglue-0.5cm
\caption{ ISGMR strength distributions in $^{208}$Pb calculated with
different parameter sets.}
\end{figure}

We are interested in establishing a link between the measured
energies of compression modes in nuclei and $K_\infty$. The most
favorable situation is met in heavy nuclei where the giant
monopole resonance is less broad than in medium and light nuclei.
In the latter case, the Landau fragmentation is quite strong due
to the weakness of the particle-hole interaction in the isoscalar
monopole channel. In order to have a comparison with
non-relativistic RPA results and with TDRMF results we concentrate
on $^{208}$Pb and $^{144}$Sm where the experimental giant monopole
energies are 14.2$\pm$0.3 and 15.4$\pm$0.3 MeV, respectively\cite{you99}.
The calculations are performed with
5 effective Lagrangians whose compression modulus span a wide
range of values: NL1 ($K_\infty =$ 211 MeV), NL3 ($K_\infty =$ 272
MeV), TM1 ($K_\infty =$ 281 MeV), NLSH ($K_\infty =$ 355 MeV), HS
($K_\infty =$ 545 MeV). The inclusion of negative energy states in
the RRPA calculations necessarily involves the choice of a cut-off
energy in the Dirac sea. The bottom of the potential on the
negative side is around -150 MeV, typically. We have checked that
the results are stable if we move the cut-off between -300 MeV and
-900 MeV.

The RRPA monopole strengths with and without Dirac states are
shown in Fig.2 for different effective Lagrangians. The strengths
are shifted down strongly to the lower energies when only the
positive energy particle-hole pairs are included. In some
parametrization, such as NL1, the strong attractive residual
interaction produced by the positive particle-hole pairs would
bring the system beyond a critical point, where the system is not
stable. In the complete RRPA calculations with the Dirac states
included the strengths are shifted to higher energies and the
unstabilities disappear. In a recent work\cite{fukui99} such an
effect due to the Dirac states has been shown analytically in the
framework of nuclear matter and the Walecka's linear model.

\begin{figure}[htb]
%\centerline{\psfig{figure=fig3.eps,height=14cm,width=14cm.angle=90}}
\epsfxsize=\hsize\epsfbox{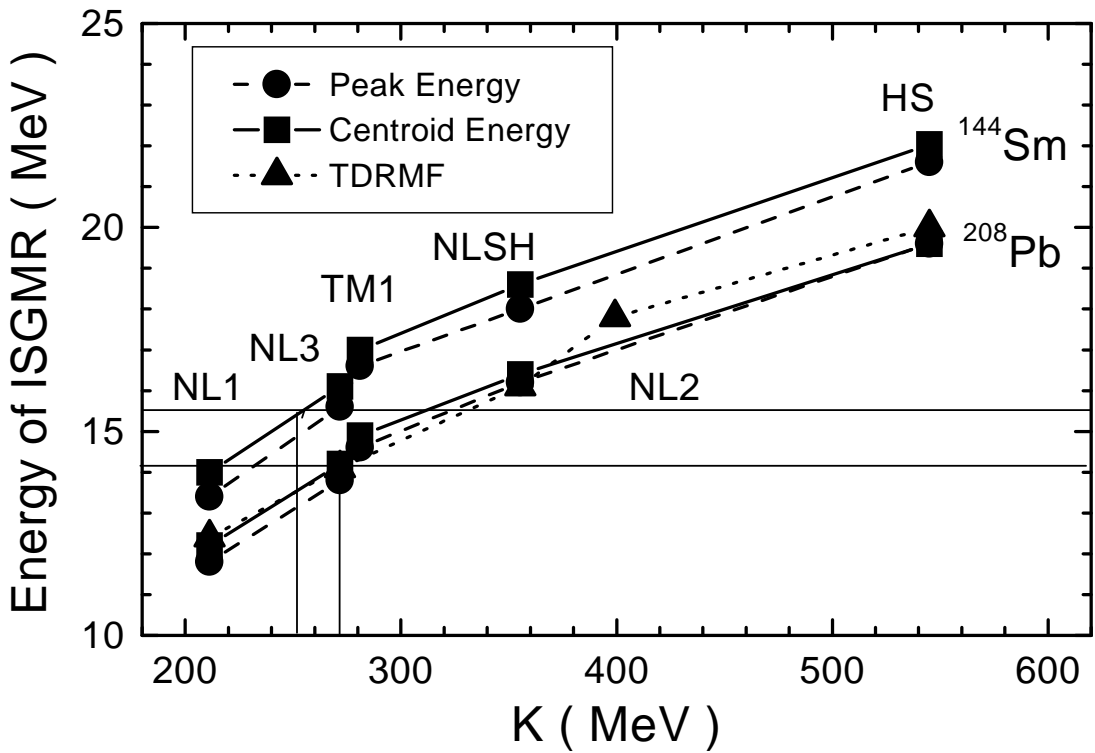}
%\vglue-0.5cm
\caption{ RRPA peaks and centroid energies of the ISGMR in
$^{208}$Pb and $^{144}$Sm as functions of $K_\infty$. TDRMF
results for $^{208}$Pb[24] are also shown. The experimental
values for $^{208}$Pb and $^{144}$Sm are
E =  14.2$\pm$0.3 and 15.4$\pm$0.3 MeV, respectively[22]. }
\end{figure}

In Fig.3 we show the RRPA monopole energies in $^{208}$Pb and
$^{144}$Sm as
functions of $K_{\infty}$. In each case, the difference between
peak and centroid energies is generally small
due to the fact that the strength is mostly concentrated in the
giant resonance region. This difference would be larger in lighter
nuclei. The ISGMR energies calculated in TDRMF\cite{vre97} are also
plotted in figure 3 in comparison with our RRPA results.
Note that the present results differ from those of Ref.\cite{gima99}
where the space-like contributions of vector mesons were not included
in the particle-hole residual interaction. A good
agreement of the ISGMR energies from the two approaches is
obtained. The dependence on $K_{\infty}$ is
nearly linear, similarly to what was obtained in non-relativistic
RPA calculations performed with Skyrme and Gogny
interactions\cite{bla95}. From Fig.3 it can be concluded that a
value of $K_{\infty}$ close to 250 $\sim$ 270 MeV would bring the RRPA
results in good agreement with experiment.

\section{Dipole compression mode}

The isoscalar giant dipole compression mode, or squeezing mode is
mainly built on 3$\hbar \omega$ excitations. It was already observed a
long time ago\cite{morsch} and it has been measured recently with
increased accuracy by ($\alpha, \alpha'$) scattering\cite{garg} in
several medium and heavy nuclei. We have performed RRPA
calculations of this mode in $^{208}$Pb.  An  external
field\cite{gia81} $\gamma_0 (r^3 - \eta r)Y_{10}$ with $\eta =
\frac {5}{3}<r^2>$ is used as a probe. The second term in the
operator is introduced to remove a large amount of spurious
components coming from the admixture of the center-of-mass state.
%included in the operator of $r^3Y_{10}$,
The value of $\eta$ is determined by a condition of translational
invariance.

It is found that there is a large amount of strength in the
1$\hbar\omega$ region around 10 MeV. The main peak (in terms of
energy-weighted strength) is at about 26 MeV for the case of NL3.
This peak moves up
with increasing incompressibility. This general pattern of
results, namely a sizable fraction of strength at lower energies
and a giant resonance peak well above 20 MeV is also found in
non-relativistic RPA calculations with Skyrme
interactions\cite{gia81,colo99}. This is at variance with the
experimental findings\cite{garg} where no strength is observed in
the 10 - 15 MeV region and where the ISGDR peak is close to 20 MeV
in $^{208}$Pb. It should be noted, however, that in the
continuum-RPA calculation of ref.\cite{ikuko97} the low-lying
strength could be eliminated by a particular subtraction procedure,
but the ISGDR peak still remains above 20 MeV.

In Fig.4 are shown the calculated peak energies as a function of
the compression modulus. The centroid energies calculated in the
giant resonance region (20-40 MeV) are also displayed. The peak
and centroid energies are generally quite close except for the
NLSH case where the giant resonance is much Landau- fragmented.
The calculated energies have a weaker $K_\infty$-dependence than in the
monopole case. The measured value\cite{garg} of centroid energy is
20.3 MeV.
One can see that a model like NL3 which is doing rather well for
the ISGMR energy is now off by 5 MeV in the ISGDR case. This
disagreement between experiment and models for the ISGDR case is
the next puzzle to be solved.

\begin{figure}[htb]
%\centerline{\psfig{figure=fig4.eps,height=14cm,width=14cm.angle=90}}
\epsfxsize=\hsize\epsfbox{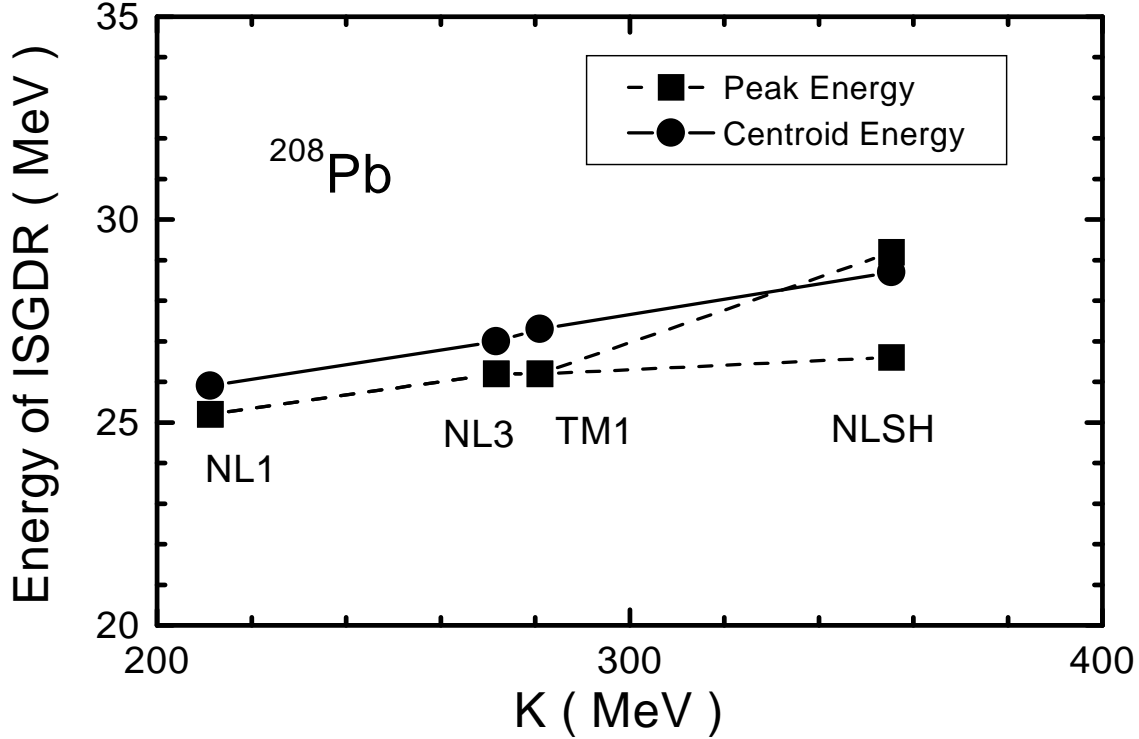}
%\vglue-0.5cm
\caption{ Peak and centroid energies (in the 20 - 40 MeV region)
of the ISGDR in $^{208}$Pb as a function of $K_\infty$. The
experimental centroid energy is E = 20.3 MeV. }
\end{figure}

\section{Summary and conclusion}

In this paper we have studied an RRPA approach built on the RMF in
the framework of the relativistic baryon-meson field theory. The
effective Lagrangians with non-linear meson self-interactions,
which describe well the ground state properties of finite nuclei
as well as the nuclear matter matter properties are adopted. The
calculation is fully consistent in the sense that the
single-particle basis and the particle-hole couplings
are obtained from the same effective Lagrangian.
In a no-sea approximation, a consistent treatment of the RRPA must
include both the
particle-hole pairs built with positive energy states and the
$\bar\alpha h$ pairs formed with the occupied Fermi sea and the
empty Dirac sea states.

  The importance of the inclusion of the Dirac sea states is examined in
detail in the case of the monopole compression mode. We have used
the constrained RMF method with a $r^2$ constraint as a check of
RRPA calculations.
With the inclusion of Dirac sea states the RRPA static
polarizability, or the inverse energy-weighted sum rule, is in
agreement with the constrained RMF result whereas the RRPA
polarizability becomes much larger (it can even be negative for
some effective Lagrangians) if the Dirac states are omitted.
In general,  the Dirac sea states must play an important role in
the isoscalar modes. The reason is that the repulsive
contributions of the $\bar{\alpha}h$ pairs through exchange of the
isoscalar-scalar meson dominate.
Their effect is to shift the isoscalar giant resonances to higher
energies.

The remaining small discrepancies between RRPA and constrained RMF
polarizabilities when the Coulomb field is turned on seem to
indicate that the harmonic oscillator basis used here may not be
fully appropriate. A proper treatment of the Coulomb interaction
in Dirac sea states for heavy nuclei is required to get accurate
results because of the long range attraction  for the negative
energy states. Employing the Green's function technique in
coordinate space and carrying out a continuum RRPA\cite{wehr} will
probably improve the accuracy of results. A good
agreement of ISGMR energies calculated in RRPA and TDRMF is
obtained when we include the Dirac state contributions in the
calculation of RRPA.

 The comparison of the isoscalar monopole data for heavy nuclei,
namely $^{208}$Pb and $^{144}$Sm, with calculations performed with
different parameter sets indicates that the compression modulus of nuclear
matter is in the range of 250 $\sim$ 270 MeV. These values are somewhat
larger than those predicted by non-relativistic RPA models\cite{bla95}.

  The main open problem is now to understand the ISGDR energies. The
discrepancies between experiment and predictions for this mode are
the largest ever found for isoscalar giant resonances described in
non-relativistic RPA or RRPA. It is unlikely that the coupling to
2p-2h configurations\cite{colo99} can account for them.

\vspace{1cm}
 \noindent
{\bf Acknowledgments}
We would like to thank U. Garg and T. Suzuki for discussions. Z.Y.M.
acknowledges the support of the SPM Department of CNRS and the
hospitality of IPN-Orsay.
The hospitality of RIKEN during completion of the work is acknowledged by
Z.Y.M. and N.V.G.
This work is partially supported by the China National Natural Science
Foundation under Grant No. 19675070, 19847002,
%This work has been supported in part
and by the Bundesministerium f\"ur Bildung und Forschung under contract
No. 06TM875. D.V. thanks the Deutsche Forschungsgemeinschaft for funding a
guest professorship at the TUM.

\begin{table}

\caption[]{The ISGMR inverse energy-weighted sum rule in
$^{208}$Pb. The rows with I,II,III refer to the results of
unperturbed (Hartree), RRPA with Dirac states and without Dirac
states, respectively. All values are in 10$^3$fm$^4$MeV$^{-1}$.}

\vspace{0.5cm}
\begin{tabular}{cccccc}
\multicolumn{2}{c}{ } &  \multicolumn{2}{c}{Constrained
RMF}& \multicolumn{2}{c}{RRPA}\\
%\cline{3-6}
\multicolumn{2}{c}{ } &
 $\frac{1}{2}\frac{\partial \langle r^2 \rangle }{\partial \lambda}$ &
 $\frac{1}{2}\frac{\partial^2 \langle E \rangle }{\partial \lambda^2}$ &
 $m_{-1}$ & $\frac{1}{2} Re \Pi(E=0)$ \\
 \hline
 & I & ~~~1.54~~~ & ~~~1.53~~~ &  ~~~1.53~~~ & ~~~1.54~~~ \\
 HS & II  & 1.48 & 1.48 & 1.51 & 1.51 \\
    & ~~~~III~~~~ & & & 2.72 & 2.73 \\
 \\
  & I & 1.66 & 1.67 & 1.66 & 1.66 \\
  ~~NL3~~ & II & 2.76 & 2.76 & 2.92 & 2.93 \\
    & III & & & 7.33 & 7.40  \\
 \\
  & I & 1.82 & 1.82 & 1.81 & 1.82 \\
  NL1 & II & 3.70 & 3.73 & 4.00 & 4.02 \\
    & III & & & 1638  & -1454 \\
 \end{tabular}
 \end{table}
\end{document}